\documentclass[preprint,12pt]{aastex}
\newcommand{\lsun}{\mbox{L$_\odot$}}% Lsun
\newcommand{\lbol}{\mbox{$L_{bol}$}} % bolometric luminosity
\newcommand{\tbol}{\mbox{$T_{bol}$}} % bolometric temperature

\input{epsf}

\def\plotfiddle#1#2#3#4#5#6#7{\centering \leavevmode
\vbox to#2{\rule{0pt}{#2}}
\includegraphics{#1}}

\begin{document}

\slugcomment{Revised Jan 18, 2009}

%%%%%%%%%%%%%%%%%% title %%%%%%%%%%%%%%%%%%%%%%%%%%%%%%%%%%%%%%%%
\title {\bf Serpens Cluster B and VV Ser Observed With High Spatial Resolution at 70, 160, and 350\micron}
\author{Paul Harvey\altaffilmark{1},
Michael M. Dunham\altaffilmark{1}
%Bruno Mer\'{\i}n\altaffilmark{2},
%Tracy L. Huard\altaffilmark{3},
%Luisa M. Rebull\altaffilmark{4},
%Nicholas Chapman\altaffilmark{5},
%Neal J. Evans II\altaffilmark{1},
%Philip C. Myers\altaffilmark{3}
}

\altaffiltext{1}{Astronomy Department, University of Texas at Austin, 1 University Station C1400, Austin, TX 78712-0259;  pmh@astro.as.utexas.edu, mdunham@astro.as.utexas.edu}
%\altaffiltext{2}{Center for Astrophysics and Space Astronomy and Department of Astrophysical and Planetary Sciences, University of Colorado, Boulder, CO 80309}

%%%%%%%%%%%%%%%%%%%% Abstract %%%%%%%%%%%%%%%%%%%%%%%%%%%%%
\begin{abstract}

We report on diffraction-limited observations in the far-infrared and sub-millimeter of
the Cluster B region of Serpens (G3-G6 Cluster) and of the Herbig Be star to the south,
VV Ser.  The observations were made with the Spitzer MIPS instrument in fine-scale mode
at 70\micron, in normal mapping mode at 160\micron\ (VV Ser only), and the CSO SHARC-II
camera at 350\micron\ (Cluster B only).  We use these data to define the spectral energy
distributions of the tightly grouped members of Cluster B, many of whose SED's peak in
the far-infrared.  We compare our results to those of the {\it c2d} survey of
Serpens and to published models for the far-infrared emission from VV Ser.
We find that values of \lbol\ and \tbol\ calculated with our new photometry show only modest changes from
previous values, and that most source SED classifications remain unchanged.

\end{abstract}

\keywords{infrared: general --- clouds: star forming regions}
%%%%%%%%%%%%%%%%%%% Main text %%%%%%%%%%%%%%%%%%%%%%%%%%%%

\section{Introduction}\label{intro}

%TODO  ????

%\noindent Make good versions of VVSer 160 figure and make YSO-locating figure

%\noindent Discuss Lbol/Tbol and c2d diffs some more

%\noindent Make multipanel SED figure
%\noindent Mike write SHARC ap flux section

%\noindent Decide on best values for uncertainties and limits (all too low now)

%\noindent Decide on division of fluxes at 350um for confused sources

%\noindent Is there any modeling worth doing?

%\noindent Add Lbol/Tbol information

The heart of the Serpens star-forming region is marked by a rich cluster of
young embedded star-forming objects that has been studied for over 30 years,
e.g. \citep{svs76, harv84, ec92}.  Roughly 3/4 of a degree to the south of this ``Core'' cluster lies
a second, somewhat less rich cluster of young objects, called ``Cluster B'' by
\citet{harv06} and named ``The G3-G6 Cluster'' by \citet{djup06}.  An additional
group of young objects in another part of Serpens has also recently been
found by \citet{guter08}.  These very young clusters of pre-main-sequence objects
contain groupings with typical separations of 10--30\arcsec, $\sim$ 0.012--0.036 pc
at the distance of 260 pc found by \citet{stra96} which we assume throughout
our study (though \citet{eiroa08} more recently find a value
of 230 pc).   Although the angular
resolution of the Spitzer Space Telescope is easily sufficient to resolve the
individual objects at $\lambda < $ 24\micron, most of the luminosity of the youngest objects
in these clusters is emitted at substantially longer wavelengths.  At 70\micron\
in Spitzer's nominal large-field survey mode used for the {\it c2d} Legacy
survey described by \citet{evans03}, the angular resolution was typically no
better than 40\arcsec\ (FWHM).  The Spitzer/MIPS instrument does, however, provide a
mode of observation that over-samples the diffraction-limited PSF of the instrument
at 70\micron\ and, at least until Herschel/PACS and SOFIA/HAWC are operational, represents
the highest angular resolution available in the far-infrared.

We, therefore, have obtained new, sensitive, diffraction-limited observations
of Cluster B in order to understand the evolutionary state of the more than one dozen
tightly clustered objects in this region.  We used the Spitzer/MIPS instrument
at 70\micron\ and the CSO/SHARC-II system at 350\micron.  The same Spitzer program also
provided diffraction-limited imaging of the Herbig Ae star VV Ser further to the south
at 70\micron\ and 160\micron\ which we discuss briefly.  We describe details of the
observations and basic data reduction in \S\ref{obs}, and then in \S\ref{anal} we describe the procedures
used to derive flux densities for
the individual sources. In \S\ref{resultsB} we then discuss the detailed results for
Cluster B and in \S\ref{resultsV} VV Ser.   In both sections we compare our
results to the earlier results from {\it c2d}.  Finally \S\ref{discuss} discusses the
effects of our improved photometry and spectral coverage on evolutionary indicators
like \lbol\ and \tbol\ as well as the SED classification.

\section{Observations and Data Reduction}\label{obs}

	\subsection{Spitzer/MIPS Observations}

Our Spitzer/MIPS observations are listed in Table \ref{aortable}.  We used the 70\micron\
fine-scale imaging in large-field mode.  For VV Ser we added 160\micron\ imaging since
the {\it c2d} maps \citep{harv07a} were not completely filled at this wavelength due to use of the
fast-scan mode with MIPS.  For Cluster B four separate areas were required to cover
fully the region of interest, and for VV Ser three fields were required.
The fine-scale imaging mode of Spitzer/MIPS at 70\micron\ provides a pixel scale that is
essentially twice as fine as the normal imaging mode.  More importantly, the pixel
scale of 5.3\arcsec\ is equivalent to $\lambda/3D$ so that the diffraction-limited PSF
is fully sampled.  Previous far-infrared studies that have also been severely
resolution limited have shown that with high S/N and good spatial sampling, it is
possible to extract some information from images at up to twice the nominal diffraction
limit, e.g. \citet{lester86, backus05, skemer08}.  We used 5 cycles of the photometry AOT for all the 70\micron\ observations
with an integration time of 3 seconds for the three AOR's on the bright part of Cluster B
and 10 seconds for the VV Ser observations and the AOR covering the faint diffuse
emission just to the northeast of Cluster B (AOR 16795904).  At 160\micron\ on VV Ser, we used 4 cycles
of the photometry AOT with 3-second frame times.

\begin{figure}
\plotfiddle{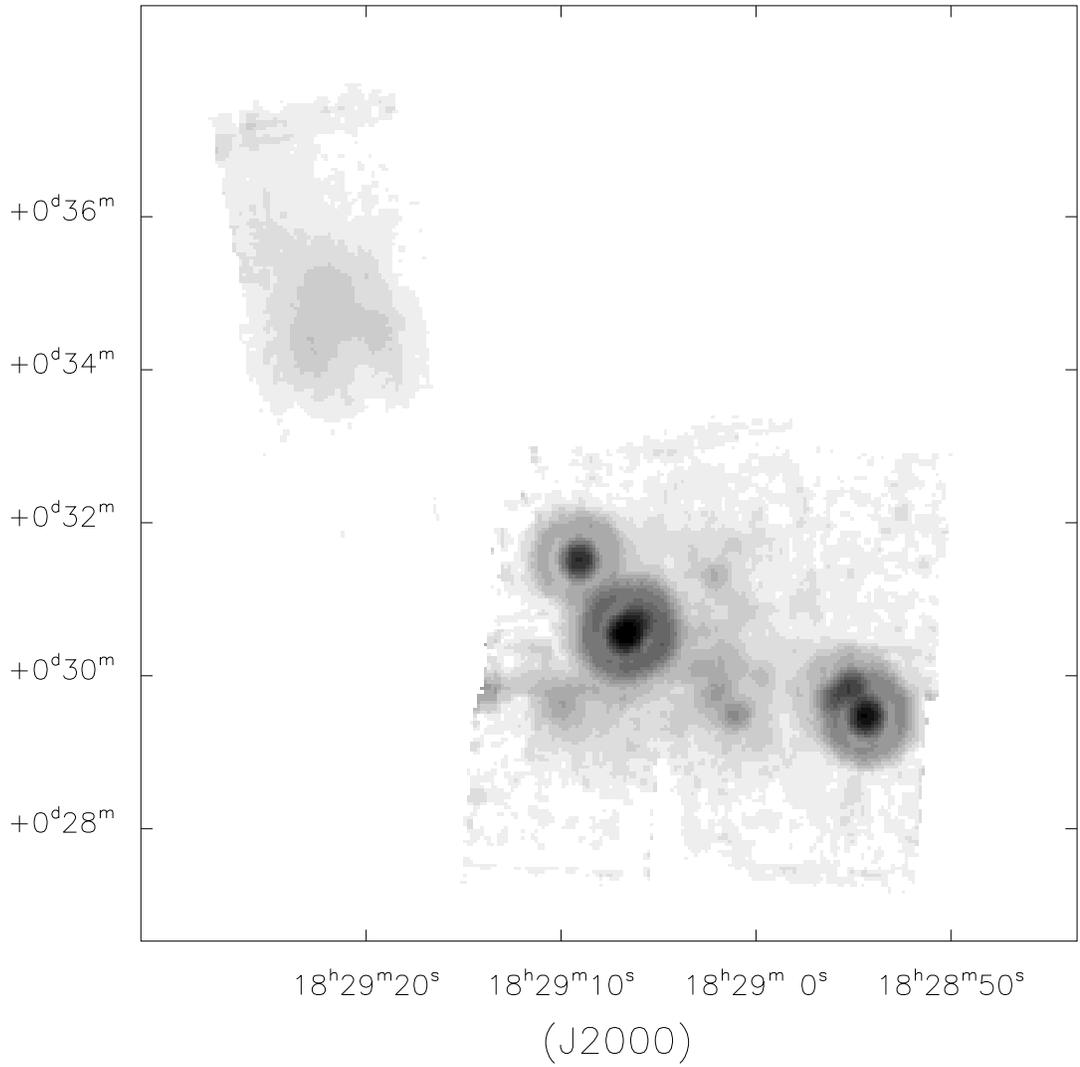}{7.0in}{0}{70}{70}{-200}{-10}
\figcaption{\label{clb70}
Spitzer MIPS fine-scale 70\micron\ image of Cluster B (Serpens G3-G6 Cluster).}
\end{figure}

The fine-scale observing mode for these observations provides alternating on-field and off-field
images with the 70\micron\ array.  This array has the most noticeable issues of any of the
MIPS arrays with problems like cosmic ray interaction and hysteresis from illumination by
bright sources or the calibration stimulator.  We initially tried mosaicking the full set
of BCD frames for each of the two fields, Cluster B and VV Ser, with parameters similar to
those used for the {\it c2d} images \citep{harv07a}.  This produced reasonably good
mosaics which, however, had several features that were cosmetically unattractive.  The
MIPS Data Handbook\footnote{http://ssc.spitzer.caltech.edu/mips/dh/mipsdatahandbook3.3.1.pdf} describes these problems and suggests several possible alternative
processing techniques to eliminate them.  One of the recommended techniques is that of
subtracting a median off-field frame from each on-field frame and then mosaicking only
the on-field frames.  Basically this technique involves producing a median of the pixel
value for each pixel in the stack of off-field frames and subtracting that median pixel value
from the same pixel in each of the on-field frames.  This technique appears to produce
cosmetically good images as shown in Figures \ref{clb70} and \ref{vvser70}, and we have therefore chosen
these images for further processing and analysis.  For the 160\micron\ data on VV Ser,
the initial mosaicking test with the full data set produced an image that appeared
relatively artifact-free, albeit with no discernible emission from VV Ser (Figure \ref{vvser160})!
The J magnitude of VV Ser is 3 magnitudes below the limit where any evidence of the
optical leak in the MIPS 160\micron\ filter would be seen as described in the MIPS Data Handbook.

\subsection{CSO/SHARC-II Observations}

Submillimeter observations of the Cluster B region of Serpens at 350 $\mu$m 
were obtained with the Submillimeter High Angular Resolution Camera II 
(SHARC-II) at the Caltech Submillimeter Observatory (CSO) on 2008 July 3.  
SHARC-II is a ``CCD-style'' bolometer array with $12 \times 32$ pixels giving 
a $2.59\arcmin \times 0.97 \arcmin$ field of view \citep{dowell03}.  
Observations can 
be conducted at 350, 450, or 850 $\mu$m by moving a filter wheel; most 
observations are conducted at 350 $\mu$m to take advantage of the 
instrument's unique ability to obtain data at this wavelength.  The 
beamsize at 350 $\mu$m is 8.5\arcsec, $\sim 1.2 \lambda/D$.

We used the box-scan observing mode to map an area approximately 10$\arcmin$ 
on a side centered at 
18$^h$ 29$^m$ 04.8$^s$ $+$00$^{\circ}\ $31\arcmin\ 10.0\arcsec, 
with a scan rate of 25$\arcsec$ sec$^{-1}$ and a spacing between adjacent 
scans of 30.3$\arcsec$.  Each scan requires 13 minutes of integration to 
fully map the $\sim 10\arcmin \times 10\arcmin$ region.  We obtained 
three scans for a total integration time of 39 minutes in moderate 
weather ($\tau_{225 \rm GHz} \sim 0.08-0.09$).  During all three 
scans the Dish Surface Optimization System (DSOS)\footnote{See 
http://www.cso.caltech.edu/dsos/DSOS\_MLeong.html} was used to correct the 
dish surface for gravitational deformations as the dish moves in elevation.

The raw scans were reduced with version 1.61 of the Comprehensive Reduction 
Utility for SHARC-II (CRUSH), a publicly available,\footnote{See 
http://www.submm.caltech.edu/\~{}sharc/crush/index.htm} Java-based software 
package.  CRUSH iteratively solves a series of models that attempt to 
reproduce the observations, taking into account both instrumental and 
atmospheric effects \citep{kovacs06} (see also Kov\'{a}cs et al. 2006; 
Beelen et al. 2006).  Pointing corrections to each scan were applied in 
reduction based on a publicly available\footnote{See 
http://www.submm.caltech.edu/\~{}sharc/analysis/pmodel/} model fit to all 
available pointing data.  We then applied an additional pointing correction 
of $+$1\arcsec\ in Right Ascension and $-$2\arcsec\ in Declination to the 
final map, based on comparison to the \emph{Spitzer} 70 $\mu$m fine-scale 
image.  The overall pointing uncertainty in the model corrections is 
$\sim 2-3\arcsec$, so this additional correction is within these 
uncertainties.

Reduced sampling near the edges of the map adds 
additional noise to these edges.  To compensate for this, we used 
\emph{imagetool}, a tool available as part of the CRUSH package, to eliminate 
the regions of the map that had a total integration time less than 25\% 
of the maximum.  We then used Starlink's \emph{stats} package to assess the 
rms noise of the map, calculated using all pixels in the off-source 
regions.  The final map shown in Figure \ref{clb350}  has a 1$\sigma$ rms noise of 190 mJy beam$^{-1}$.
Figure \ref{clb2470350} shows a false color composite image of the Spitzer 24 and 70\micron\ images
(blue and green) and the SHARC-II 350\micron\ image (red) in the area of overlap.
For clarity, we also show the contours of the 350\micron\ emission superimposed
on the image.

\section{Determination of Flux Densities}\label{anal}

\subsection{Spitzer 70\micron\ Fluxes}

Within the observed field of Cluster B in this study there are 17 young
stellar objects (YSOs) from the study by \citet{harv07b} plus two
very embedded objects (B and C) from Table 7 in \citet{harv07a} that did not
make the stringent cut for YSOs because of the lack of detectable 3.6\micron\
emission.  Table \ref{ysotable} lists all 19 objects with their YSO number
for the 17 from \citet{harv07b} and with the SST designation which includes their
RA and Dec.  In the remainder of this study we will refer to each source by
its YSO number (or letter) in Table \ref{ysotable}.  The locations of these sources
in our new 70\micron\ imaging are shown in Figure \ref{ysolabel}.  
No point sources are visible in our new images that are not
in this list, and several in the list are, in fact, undetected even in our
new images that have better sensitivity and angular resolution than the {\it c2d} images.
There are, however, two regions of extended emission seen
at 18$^h$ 29$^m$ 10$^s$ $+00^{\circ}\ 29\arcmin\ 40\arcsec$  and 18$^h$ 29$^m$ 22$^s$ $+00^{\circ}\ 34\arcmin\ 40\arcsec$.  
The latter was specifically included in our observations
as a separate AOR (Figure \ref{clb70}) to see if improved 70\micron\ imaging might
clarify the nature of this region which includes diffuse emission at
most Spitzer wavelengths. No compact source of excitation is visible in our
new data.  These diffuse emission regions will not be
discussed further in this paper.  In the 350\micron\ image (Figure \ref{clb350}) it
is also true that no compact sources are seen that are not associated with
one of the objects in Table \ref{ysotable}.

Since the goal of this study was to obtain the highest quality photometry
possible at 70\micron, we explored several methods to estimate the
flux densities in this crowded cluster.  
The original  {\it c2d} 70\micron\ flux densities were derived using the
SSC's Apex source extractor for historical reasons, but in general we have
had more experience with the internal  {\it c2d} source extractor, c2dphot.
This software is based on the
operation of Dophot \citep{dophot}.  Specifically, the software searches
for peaks at increasingly lower flux levels; when a peak is found with
sufficient S/N, it is fit with the PSF (or an extended ellipse if necessary) 
and subtracted from the image.  This works very well in most of the {\it c2d} fields
as described in the Delivery Document for the project \citep{evans07}.
Therefore, we tried running c2dphot on our final mosaic as a first test.
Three sources were undetectable by eye and by the source extractor (\#'s 37, 41, and 46).
Seven of the 19 sources in Table \ref{ysotable} were not extracted 
because of their close proximity to nearby brighter sources, even though by
eye it is possible to distinguish several of them (\#'s 42, 45, 56, 59, 67, 75, and B). This suggested that the extracted
fluxes for the other sources might also have problems as well.   For example,
once a brighter source is characterized and subtracted, small differences between
the assumed and true PSF's will lead to larger effects on fainter nearby
objects.

We, therefore, decided to investigate an algorithm that could simultaneously
fit all 19 point sources that might be in the image, rather than fitting
them sequentially and subtracting each after it was fitted.  In order to
maximize the quality of the flux determinations, we chose to keep the
source positions fixed in the fitting process.  Since all the
visible compact sources appear by eye to be coincident with their shorter
wavelength counterparts within the mutual positional uncertainties, this
does not represent a significant compromise.  Furthermore, at the end
of the process, we have subtracted the final estimated contributions of
each fitted object from the image to check the reasonableness of this
process as described later and shown in Figure \ref{clb70sub}.  For our model fitting we chose the
``amoeba'' function \citep{nr} which is an implementation of the
simplex algorithm.  The free parameters were the 19 flux densities of the
known, shorter wavelength point sources with strong 24\micron\ emission
plus two parameters to represent a constant background level with an east-west gradient across
the image that appeared present at a low level.  
%Although the assumption of a constant background is not completely consistent with
%the image, the background level in the image appears to be similar within
%10\% in the four corners.   
We ran the algorithm several times with
increasingly restrictive tolerances on the allowed change in fit to
the  $\chi^2$ values and watched how the fitted flux densities varied.
The final tolerance level was $1 \times 10^{-5}$ for the maximum allowed change
in  $\chi^2$ per iteration.  In general, the sources
that were reasonably isolated showed little variation in different runs,
but the objects that were partially confused with nearby sources showed
an unpleasantly large variation in several cases.  We estimated uncertainties
in the fitted flux densities for the relatively isolated sources
by re-running the model fit with the flux of
each object independently held fixed at levels above and below the best-fit
value and noting the increase in $\chi^2$.  The uncertainties were
calculated as the change in flux necessary to produce a $\Delta \chi^2 = 1$.
The background level and its slope were typically constrained at the 5-10\% level,
and none of the sources detected above 100 mJy was affected by these uncertainties
n background at greater than the 5\% level.

In order to increase our confidence in the extracted fluxes for the
highly confused objects, we added one more stage of analysis to this
process.  For the simple double sources where the ratio of flux densities was typically 
between 1:1 and 3:1 (\#'s 42/45, 54/56, 58/59, and 67/68), we
explored the $\chi^2$ space for a large range of flux values for each
of the two sources while holding the fluxes for the other 17 objects
(and the background) fixed at the best value determined from the simplex
fit to all of them.   Figure \ref{chisq} shows an example of the result from
this exploration for one of the four cases; all produced similar results.
As expected, there is a strong correlation between the fluxes of the two
objects in the sense that their sum is more strongly constrained than
either one individually.  With these results, then, we can choose the
highest likelihood estimate as well as the uncertainties in these
estimates as described by \citet{nr}.  The uncertainties are
marked by the extent of the $\chi^2$ contours equal to 2.3, the value
appropriate for two free parameters as shown in the figure.

For the case of the three sources in northeast corner of the map,
\# 75, B, and C, we tried several methods to separate the fainter
objects from the bright object, C, that exhibits the ``coldest'' SED.
No detectable emission with reasonable S/N was found either by
treating the group as 3 pairs or by running the amoeba function
on this group alone.  Careful visual inspection of the image also showed
no reliable evidence for emission from either of the fainter objects
in the wings of source C.  Therefore, Table \ref{ysotable} gives
upper limits for these two sources derived from the range of  $\chi^2$
found in our fitting attempts for these sources.
%UNC for isol sources

Table \ref{ysotable} lists the final flux densities found and the
total uncertainties including some additional factors discussed
below.  While producing the image with the point sources subtracted
(Figure \ref{clb70sub}) we realized that the region around sources 40, 42, and 45
is particularly sensitive to the near wings of the assumed PSF.  In the first
subtracted image we produced, there was a bright area in the region between all
three sources (i.e. $\sim$ 8\arcsec NE of source 40), that we realized was due to
the under-subtraction of the three overlapping source wings there.  We used
this fact and other aspects of the subtracted image to attempt to refine the assumed PSF.  
Roughly speaking, it was clear that the assumed PSF was slightly too narrow and the height
of the first bright diffraction ring in the assumed PSF was too low.  We experimented
with the Spitzer TinyTim PSF tool\footnote{http://ssc.spitzer.caltech.edu/archanaly/contributed/stinytim/index.html}
and ended up with a PSF that is shown in Figure
\ref{psffigure}.  We created this somewhat artificial but well-fitting function by
running TinyTim with 2.5\arcsec\ of jitter for a source with color temperature of
50 K (close to the coldest of our sources).  In order to fit the first bright diffraction
ring, we added a low-level Gaussian with a half-width of 10 pixels (FWHM) and peak height
5\% of the TinyTim PSF.  Figure \ref{psffigure}
shows that there are still some small differences, even in this 1-D cut, but the overall
agreement is rather good. 

This was the PSF used to derive the final fluxes.  
%It is clear,
%however, from the subtracted image that there are still obvious differences between
%the true PSF and our slightly smoothed version of the one from the SSC website (REF).
%All of the subtracted images of the individual objects show a negative minimum in
%the center and a positive (above the background) ring.  This suggests that the true
%PSF is slightly broader in the more distant wings than even our slightly smoothed
%version.  
The remaining small differences between the assumed and true PSF's could possibly
lead to small errors in the flux estimation. In principle, though, we can
check our calibration by comparing our derived flux densities with
a simple addition of surface brightness in the map for individual sources.  This is, in fact,
how the original c2dphot PSF-fitted fluxes were calibrated.  In our mapped area, there is no
single, isolated bright source; there is, however, the grouping of three sources
mentioned above that indicated the small PSF problem, sources 40, 42, and 45.  The
nearest other sources are substantially fainter.  So we have summed the total
surface brightness in the map around this cluster and multiplied by the pixel
solid angle to obtain a total of 21.3 Jy.  In Table \ref{ysotable} we can see
that the modeled total for these three sources is 21.6 Jy.  A second, less accurate
test can be made with sources 75, B, and C, where we have tried to separate their
contribution from the bright pair, 67 and 68.  In this case we find the flux
sum from the image implies a total for these of 9.4 Jy while the total from
Table \ref{ysotable} is 8.7 Jy.  Given the difficulty of separating the two groupings,
we consider this consistent with the result for sources 40, 42, and 45.  Therefore,
we believe the flux uncertainties associated with the extraction process itself
are less than 10\% except for the confused triple around source C, and we
have assigned flux uncertainties of $\pm$ 10\% for all the fluxes that had smaller
formal uncertainties.  Tests done comparing TinyTim PSF's for 50K and 400K blackbody
sources suggest that the error due to assigning a single color-temperature PSF for
all our sources is less than 5--6\% for the warmest few detected sources.
In addition, the MIPS data handbook and the description of the absolute calibration
process \citep{mipscal} suggest there is an absolute calibration uncertainty of order
5\% due to repeatability of the MIPS calibration observations.  We have therefore
assigned a minimum absolute flux uncertainty in Table \ref{ysotable} of 20\% to include
errors in both our flux extraction process, absolute calibration, and any other
subtle systematic errors.

%	\subsection{Spitzer Images}

%	\subsection{SHARC Images}

%	\subsection{Flux Estimation}
%\subsection{SHARC 350\micron\ Fluxes}

%The situation for determining the fluxes of the sources detected at 350\micron\ is
%more problematic because only some appear to be well-fitted by the SHARC PSF.

\subsection{SHARC-II 350 $\mu$m Fluxes}

The situation for determining the fluxes of the sources detected at 350 
$\mu$m is more problematic because only some appear to be well-fitted by 
the SHARC-II PSF, determined from observations of several calibration sources.
In particular, we tried the same technique described above for PSF-fitting and found
from the subtracted image
that the close double source (\#'s 67 and 68) could not be fit at all as the sum
of two PSF's, even allowing for some positional mismatch between the SHARC-II and Spitzer
observations.  The same was true for source 40 which was also the one shown in Figure
\ref{clb70sub} with the worst fit at 70\micron\ in the subtracted image. 
This is, of course, not surprising since the 350\micron\ emission traces quite cool
dust that is likely to be far enough from the central stars to be spatially extended.
A second problem for the complex of
three sources in the northeast, 75, B, and C, is that very small differences in
assumed position registration between the SHARC-II map and the Spitzer map lead to
enormous changes in the derived flux densities for the two faint objects
near the brightest component of the triple.  Therefore, we have decided to use
aperture fluxes at 350\micron\ and make some attempt to divide the total fluxes
for confused objects between the individual objects.  We show larger uncertainties
for these.

A total of six sources are detected in the SHARC-II map 
(Sources 40, 42, 45, 67, 68, and C).  
We calculated flux densities in 20\arcsec\ and 40\arcsec\ diameter apertures, 
centered at the peak positions of the sources, for each source detected.  
The method, based on the requirement that a point source should 
have the same flux density in all apertures with diameters greater 
than the beam FWHM (8.5\arcsec\ for these observations), 
is described by Wu et al. (2007) (see also Shirley et al. 2000) 
and briefly summarized here.  
Flux conversion factors (FCFs) 
are calculated for each aperture by dividing the total flux density of a 
calibration source in Jy by the calculated flux density 
in the native instrument 
units of $\mu$V in each aperture.  Flux densities 
of science targets are then derived by multiplying the aperture flux density 
(in the instrument units) of the source by the appropriate FCF.

We chose 20\arcsec\ diameter apertures 
as the most reliable estimates of total flux density for 
Sources 40, 42, and 45; such apertures represent the best  
compromise between including the full extent of the source emission and 
avoiding overlap with neighboring sources.  We chose 40\arcsec\ diameter 
apertures for the sum of Sources 67 and 68 and for Source C.
It was impossible to obtain a reliable model fit to the 350\micron\ fluxes
for Sources 67 and 68, because of the positional uncertainties.  The
image clearly shows an extended, roughly elliptical, source roughly centered
on the position between the two YSO's.  Unpublished CARMA 3mm interferometry of
this region \citep{enoch09} shows two sources of roughly equal
flux at the locations of the two Spitzer sources.  We have therefore divided the total
flux for the pair equally between each one and assigned uncertainties of $\pm$45\%
to these estimates.

The 350 $\mu$m 3$\sigma$ upper limit for all sources not detected is 
600 mJy, except for Source 75 where its close proximity to Source C leads
to a much higher limit, estimated as 3 Jy.

\section{Results for Cluster B and Comparison to {\it c2d}}\label{resultsB}

%	\subsection{Comparison With {\it c2d} Results}

Table \ref{ysotable} lists the 70\micron\ flux densities for the objects detected in
the {\it c2d} survey and reported by \citet{harv07a} and \citet{harv07b} along with the flux
densities determined above from our new, higher resolution data.  There is clearly
very poor agreement within the stated uncertainties for most of these flux densities.
By far the majority of the most significant discrepancies for the higher S/N detections
are in the sense of our measurements finding a larger value than did {\it c2d}.  The
most likely reason for this is that the {\it c2d} measurements were much more strongly
affected by non-linearity or possibly saturation effects than our new fine-pixel-scale data.
The {\it c2d} data were taken in MIPS fast-scan mode which has an integration time
per frame of 3 sec.   The MIPS Observers Manual\footnote{http://ssc.spitzer.caltech.edu/documents/SOM/som8.0.mips.pdf}
suggests that the saturation
limit for the MIPS 70\micron\ channel in wide-field mode is 23 Jy per second of
integration, i.e. 7--8 Jy for 3 second integration times. 
Although the MIPS pipeline attempts to correct for mildly saturated sources by
using only the first few ramp samples, there may still be uncorrected non-linearities
at these flux levels.
% This would imply saturation limits of order 7-8 Jy.  
This is certainly
consistent with the comparison seen in Table \ref{ysotable} where the largest
flux discrepancies are for the cases where our new measurements are at levels of
8 Jy and above.  For example, even in the case of source C, the deeply embedded
object described by \citet{harv07a}, 
the sum of the fluxes of the three sources in our new data, 8.7 Jy, should be compared with
%it was not separated from the nearby sources 75 and B, so that the sum of their fluxes of 8.7 Jy is the value to be compared with
the {\it c2d} value of 6.4 Jy.  The most striking example of a {\it weaker} flux measurement
in our new data is for source 42 which is one of the two fainter members in the cluster
of 3 objects in the southwest corner of our map.  It seems likely that the previous
measurement was confused by the presence of the much brighter source 40 located
23\arcsec ($\sim1.5 \lambda/D$) to the southwest.

Another aspect of this comparison is to examine the sources that were not extracted in
the original {\it c2d} maps and to consider to what extent our new measurements can
be considered reliable.  These are the sources in Table \ref{ysotable} without
70\micron\ {\it c2d} fluxes.  Five of these, \#'s 37, 41, 46, 55, and 56 were also
not detected in our new data set, and are clearly not detectable by eye in either
the {\it c2d} image or our newer, higher-resolution imaging.  That leaves two
objects, \#'s 45 and 58, that were detected reliably only with our new observations.
Source 45 is clearly visible by eye in our new image and also is barely visible
in the {\it c2d} 70\micron\ image.  It was not extracted in the {\it c2d} processing
because of its close proximity to the much brighter source 40.  Source 58 is faint
and closely blended with the nearby faint source 59, and the two cannot be readily
distinguished from a single object in the image within the S/N.  The excellent
degree of subtraction in this region of the image shown in Figure \ref{clb70sub}
suggests, though, that our estimation of the fluxes of the two sources (clearly
visible separately at 24\micron) is probably reasonably accurate.  Figure \ref{chisq}
also suggests that the division of fluxes between these two sources is fairly reliable.

\section{Results for VV Ser}\label{resultsV}

\begin{figure}
\plotfiddle{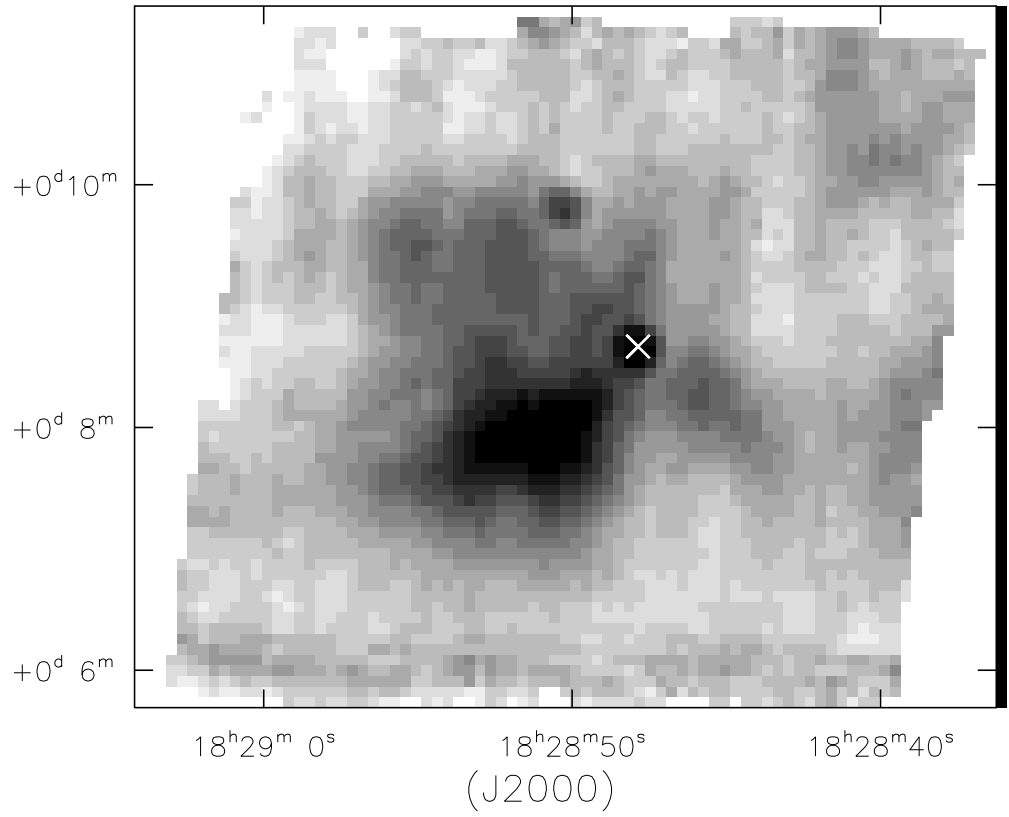}{7.0in}{0}{70}{70}{-200}{-10}
\figcaption{\label{vvser70}
Spitzer MIPS fine-scale 70\micron\ image of VV Ser. The nominal position of VV Ser is
marked with the white 'X'.}
\end{figure}

\begin{figure}
\plotfiddle{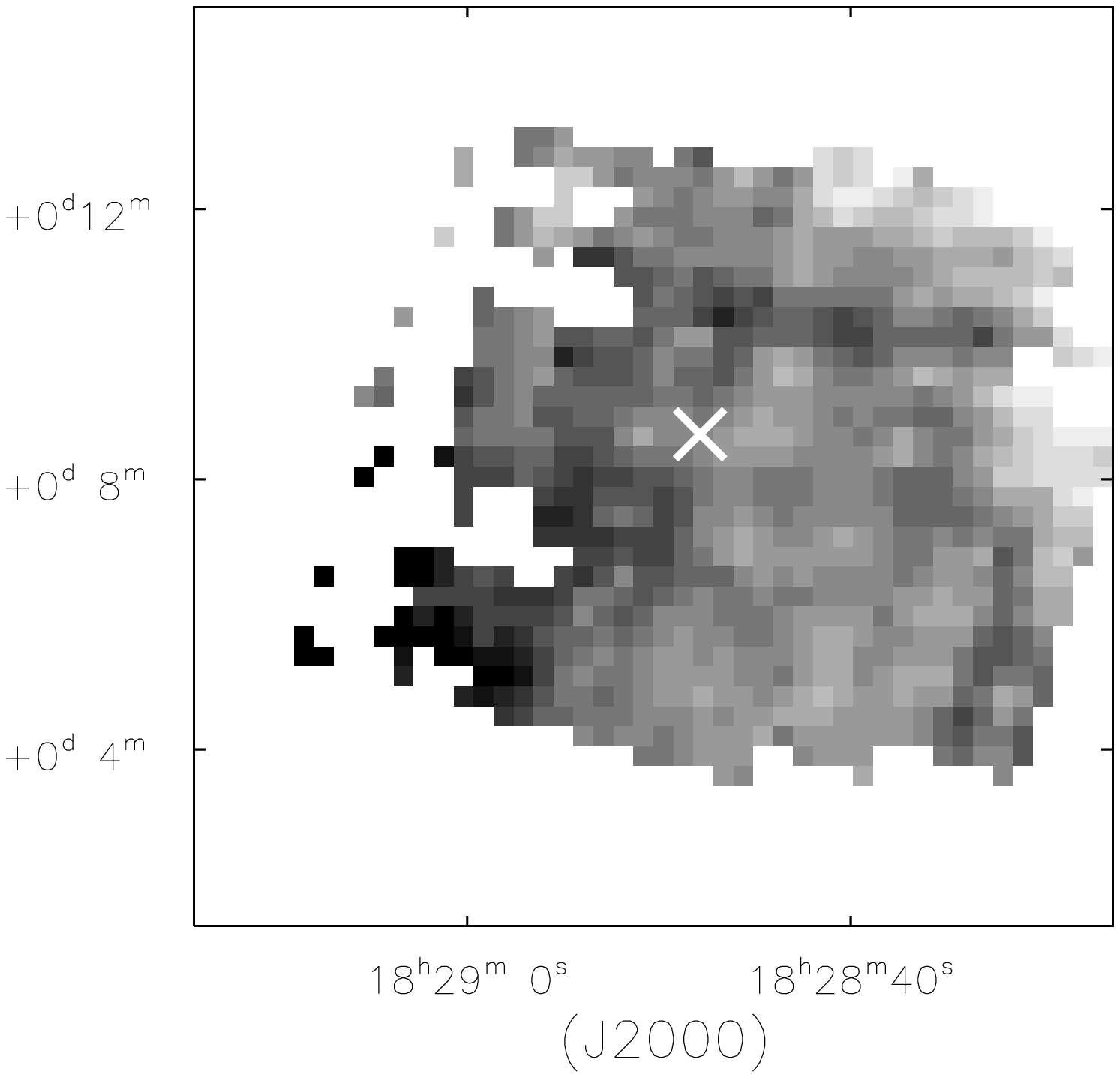}{7.0in}{0}{80}{80}{-280}{-10}
\figcaption{\label{vvser160}
Spitzer MIPS 160\micron\ image of VV Ser. The nominal position of VV Ser is
marked with the white 'X'.}
\end{figure}

\begin{figure}
\plotfiddle{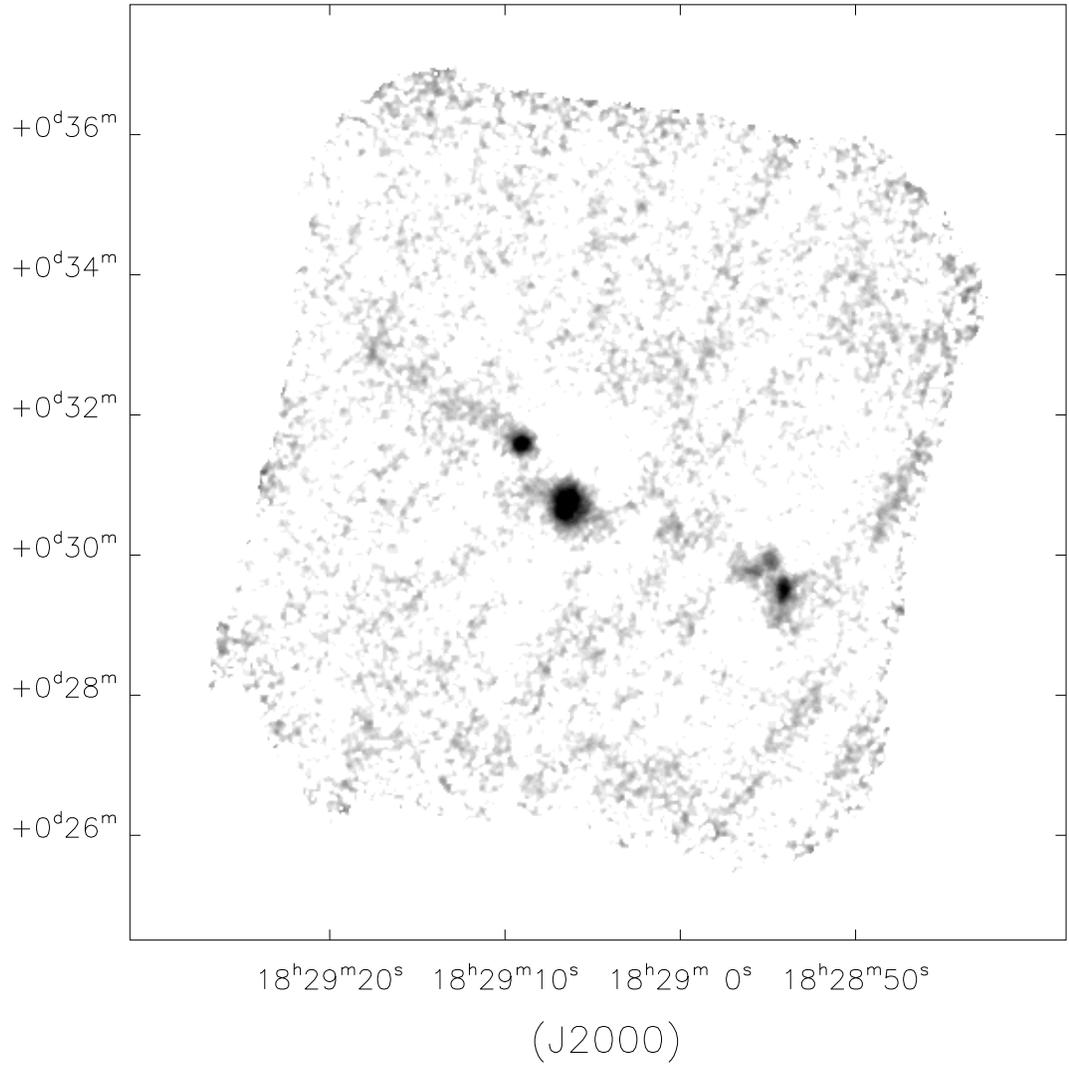}{7.0in}{0}{70}{70}{-200}{-10}
\figcaption{\label{clb350}
CSO SHARC-II 350\micron\ image of Cluster B.}
\end{figure}

\begin{figure}
\plotfiddle{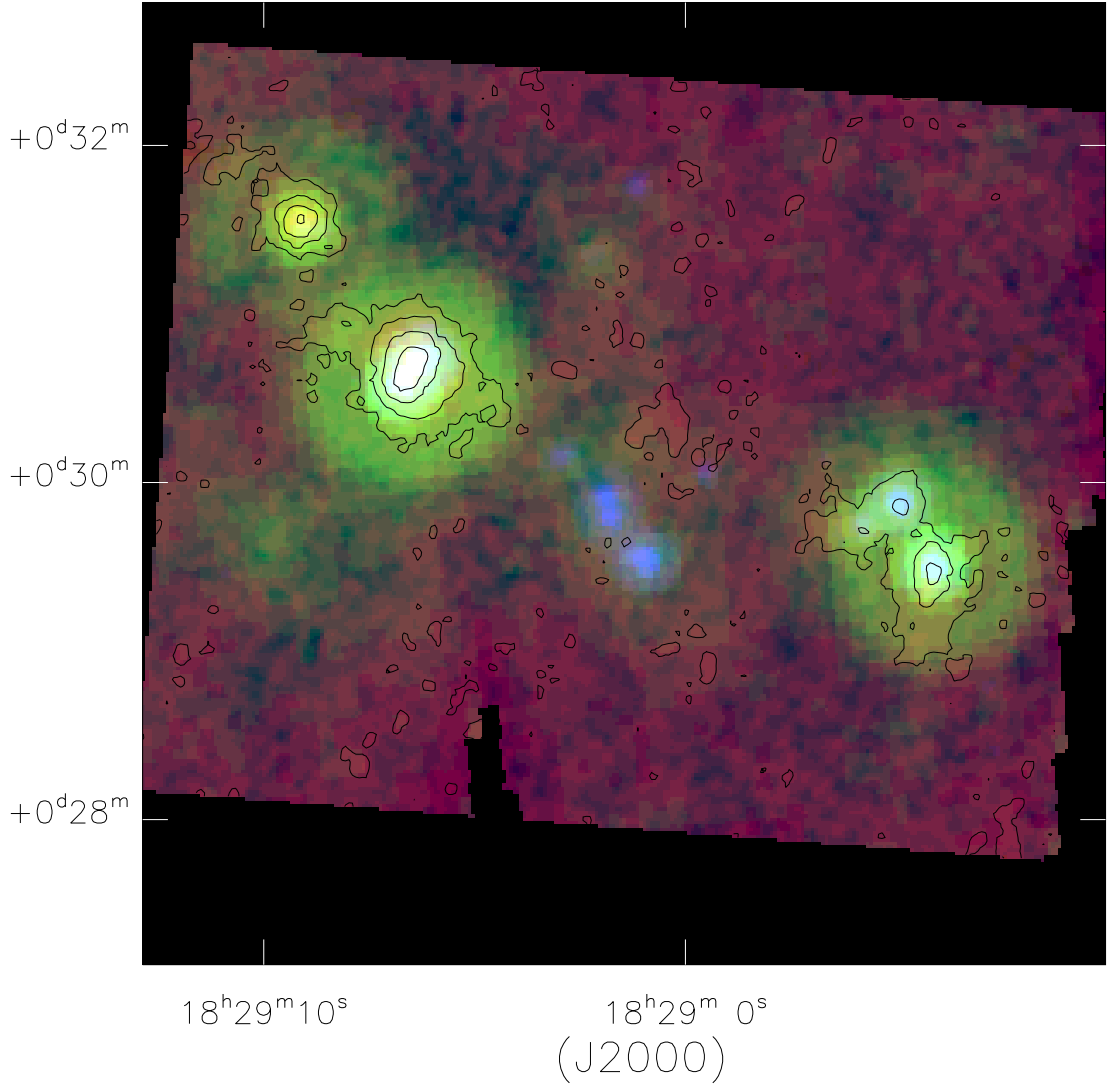}{7.0in}{90}{80}{80}{350}{-10}
\figcaption{\label{clb2470350}
False color image of Cluster B with the 24\micron\ image from the {\it c2d} data (blue),
the 70\micron\ fine-scale image from this study (green), and the SHARC-II 350\micron\ image from
this study (red).  The contours of the 350\micron\ data are overlaid on the image for clarity.
The contours are logarithmically spaced at levels of $Log_{10}$ Surface Brightness (MJy/sr) = 2.5, 2.9, 3.3, and 3.7.
}
\end{figure}

\begin{figure}
\plotfiddle{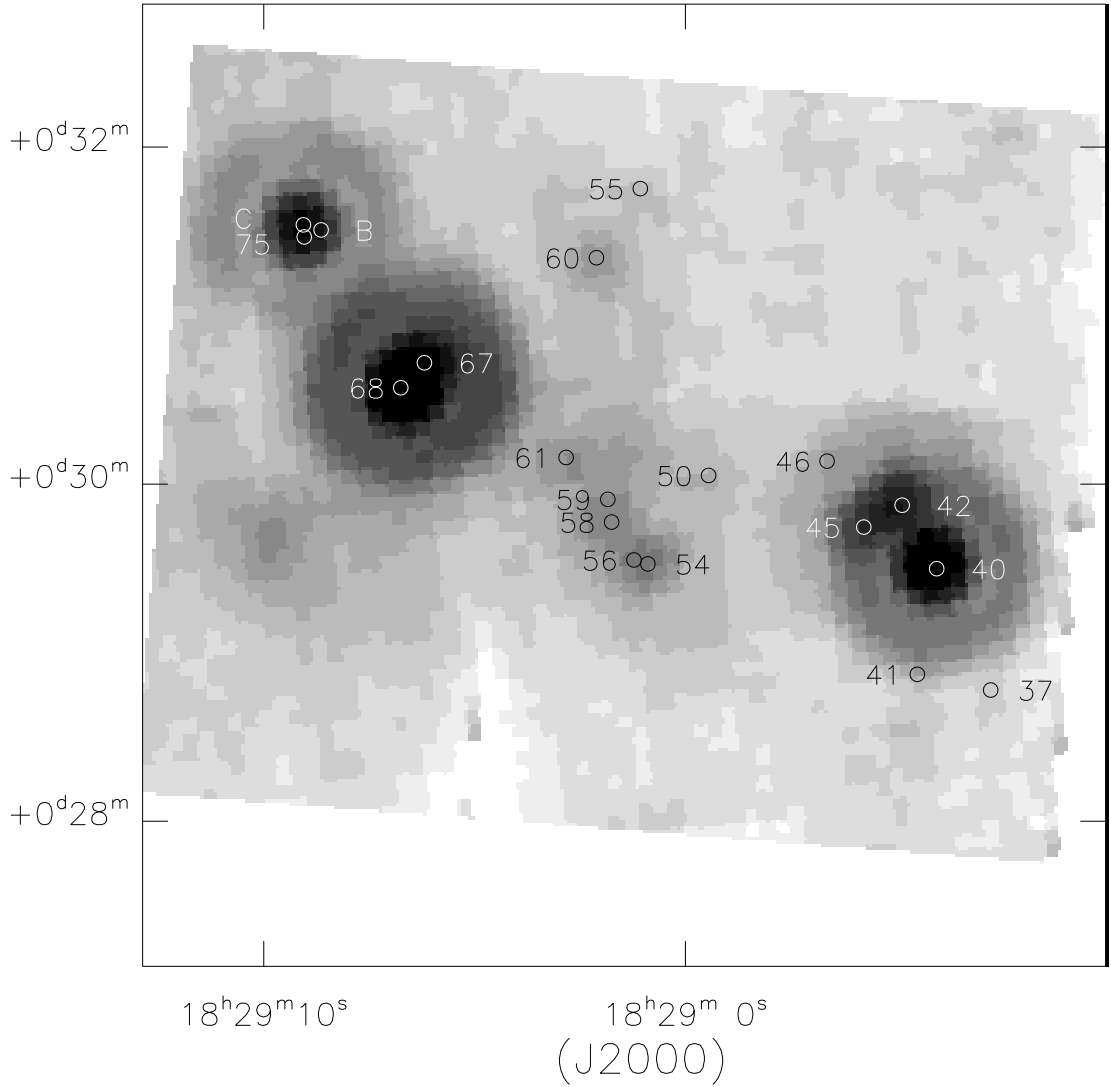}{7.0in}{90}{80}{80}{350}{-10}
\figcaption{\label{ysolabel}
Image of the Cluster B region from figure \ref{clb70} with the locations of the
19 YSO's or candidates from Table \ref{ysotable} marked.
}
\end{figure}

\begin{figure}
\plotfiddle{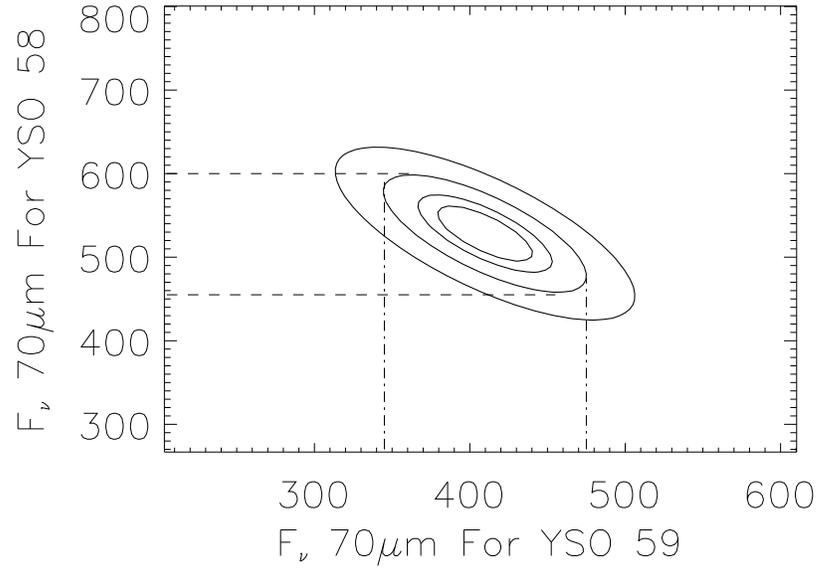}{7.0in}{0}{70}{70}{-200}{-10}
\figcaption{\label{chisq}
Contours of the $\chi^2$ values of the model fit for two closely spaced
sources.  The contours are at values of $\chi^2$ above the best-fit value by
0.5, 1.0, 2.3, and 5.0.  Lines are drawn indicating the $\chi^2 = 2.3$ limits
above the best fit value delimiting the uncertainties. 
}
\end{figure}

\begin{figure}
\plotfiddle{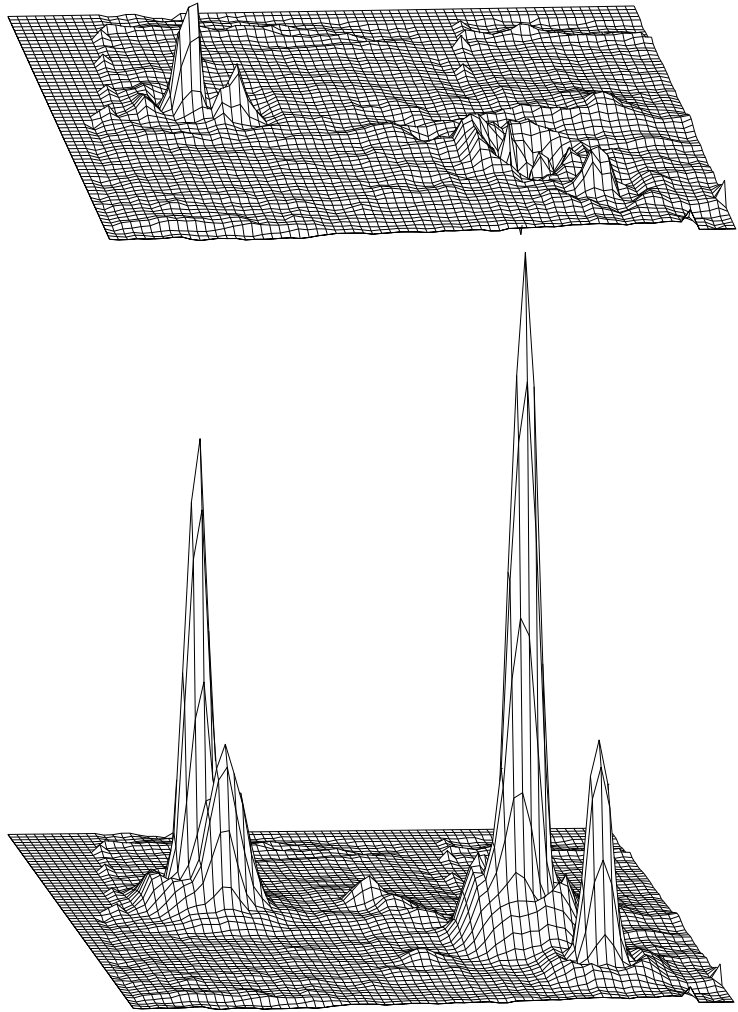}{7.0in}{0}{70}{70}{-200}{-10}
%\plotfiddle{ds9figsubalt.ps}{7.0in}{0}{70}{70}{-200}{-10}
\figcaption{\label{clb70sub}
Lower panel: surface plot of our 70\micron\ image of Cluster B, viewed
roughly from the north.  Upper panel: image with the point sources
subtracted as described in the text.  The maximum flux level is
identical in both panels and is 2000 MJy/sr.  Most of the residuals are well
under 10\% of the original image with the exception of some emission
in the vicinity of sources 40, 42, and 45 that may be due to a slightly
incorrect estimation of the PSF or to actual weak extended emission.}
%Left panel: Spitzer MIPS 70\micron\ image of Cluster B with positions of
%sources from Table \ref{ysotable} marked; Right panel: Same image with fitted
%point sources subtracted from image.}
\end{figure}

\begin{figure}
\plotfiddle{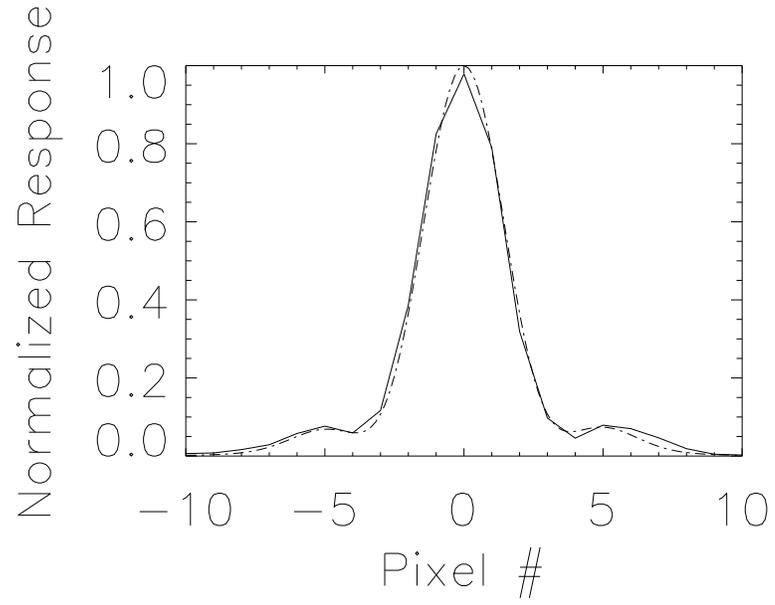}{7.0in}{0}{70}{70}{-200}{-10}
\figcaption{\label{psffigure}
Plot of a 1-D cut through source C along a roughly east-west
line (solid) versus the final assumed PSF (dash-dot).
}
\end{figure}

The Herbig Be star VV Ser is also a member of the UX Ori class whose members are believed
to be surrounded by nearly edge-on disks \citep{pont07a}.  We included this
object in our study in order to search for possible structure in the surrounding
nebula that has been modeled by \citet{pont07b} as well as to obtain a better
flux density measurement at 160\micron\ than the {\it c2d} data that were
not fully sampled spatially as mentioned earlier.  Our 70\micron\ fine-scale
image (Figure \ref{vvser70}) is qualitatively and quantitatively quite similar
to that from the {\it c2d} dataset shown by \citet{pont07b}, though the derived flux
differs as discussed below because of the very preliminary analysis used by  \citet{pont07b}.  We certainly have
not identified any small scale structure within the nebula.  Our 160\micron\
image is also quantitatively consistent with the {\it c2d} dataset in that we
see no obvious evidence for a compact emission source associated with the
star.  If anything, there is a slightly lower level of emission in the center
of the image where VV Ser is located than at the eastern edge of the image.

We can derive a limit on the 160\micron\ flux density of 4 Jy
from our image and a new measurement of the 70\micron\ flux density from
that in Figure \ref{vvser70} of 630 mJy.  This 70\micron\ flux density is
essentially identical to that in the final {\it c2d} data delivery, but is
nearly a factor of 2 greater than that used in the modeling by \citet{pont07a} and \citet{pont07b},
because they were working from a preliminary analysis of the {\it c2d} data.
These values are quite consistent, however, with their models,
because their model flux density for the star at 70\micron\ was $\sim$ 600 mJy and
at 160\micron\ was less than 100 mJy.  Our flux densities are more than a factor of
10 fainter than the values found by IRAS at 60 and 100\micron\ as illustrated in
Figure 3 of \citet{pont07b}.  This is almost certainly because of the much larger
beam size of IRAS together with the extensive and structured diffuse emission
in the region as easily seen in the images of \citet{harv07a}.

\section{Discussion}\label{discuss}

	\subsection{Spectral Energy Distributions}

\begin{figure}
\plotfiddle{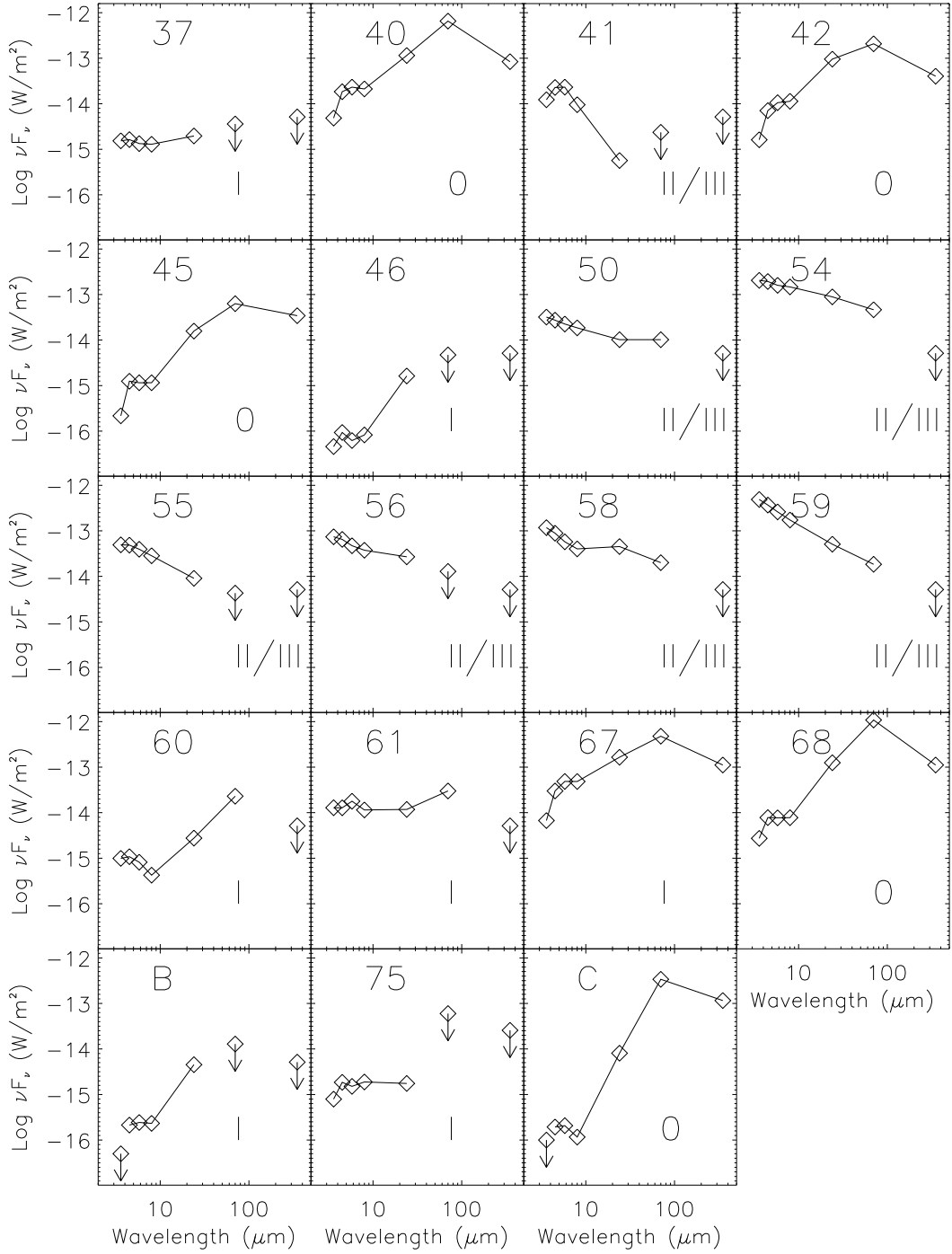}{7.0in}{0}{70}{70}{-200}{-10}
\figcaption{\label{sed}
Spectral energy distributions of the Cluster B YSO's and candidates in
Table \ref{ysotable} with the SED classification from Table \ref{tab_lboltbol} shown as well.
}
\end{figure}

Figure \ref{sed} displays the SED's for all the sources in the mapped area of
Cluster B. This figure shows a huge variety of SED's among the members of Cluster B,
a fact that was already noted by \citet{harv07b} and  in the Perseus Cloud by \citet{rebull07}.
As both these studies discussed, the range of evolutionary states implied by this mix
of SED classes (and bolometric temperatures discussed below) suggests that the members of this 
cluster probably began their lives within some range of formation times and/or have evolved
at different rates since then.

\subsection{Bolometric Luminosities and Temperatures}

We calculate the bolometric luminosity (\lbol) and bolometric temperature 
(\tbol; Myers \& Ladd 1993) for all 19 YSOs using the flux densities 
presented in Table 2, 2MASS flux densities (if detected), 1.3 mm 
flux densities from Djupvik et al. (2006), and 160 $\mu$m and 1.1 mm 
flux densities from \emph{Spitzer} and Bolocam compiled by 
Evans et al. (2008) 
(see Enoch et al. 2007 for the original Bolocam study).  
Where both 1.1 and 1.3 mm flux densities are available, we use only the 1.3 
mm results since the beamsize of these observations was smaller 
(11\arcsec\ FWHM vs. 30\arcsec\ FWHM).  The integration over the 
finitely sampled source SEDs is done using the trapezoid method.  Our results 
are presented in Table \ref{tab_lboltbol}.  We do not list uncertainties in 
either quantity; the error introduced by integrating over incomplete, 
finitely-sampled SEDs is typically $20-60$\% 
(Enoch et al. 2008; Dunham et al. 2008), larger than the statistical 
uncertainties from propagation of the photometric uncertainties (typically 
$\sim$ 10\%).

Table \ref{tab_lboltbol} also lists the classification of each source 
based on the \tbol\ classification scheme of Chen et al. (1995).  Since 
the photometry used to calculate \tbol\ is uncorrected for extinction, 
this classification method does not distinguish between Class II and III 
objects (Evans et al. 2008), 
thus we list all objects with \tbol\ $\geq$ 650 K, the dividing 
line between Class I and Class II according to Chen et al. (1995), 
as Class II/III.  Also, luminosities for Class II/III objects are 
best treated as lower limits since no extinction corrections are applied.

\subsection{Effects of Improved Photometry on Evolutionary Indicators}

The last two columns of Table \ref{tab_lboltbol} list the values of 
\lbol\ and \tbol\ for the same sample of sources calculated by Evans 
et al. (2008), using the same data except default-scale 70 $\mu$m 
images rather than fine-scale and no 350 $\mu$m photometry.  Even with 
the improved photometry available through this study, only one source 
changes classification (Source 68 changes from Class I to Class 0; 
also note that Source 60 moves very close to the Class 0/I boundary of 
\tbol\ $=70$ K).

\begin{figure}[t]
%\plotone{f11.eps}
\plotfiddle{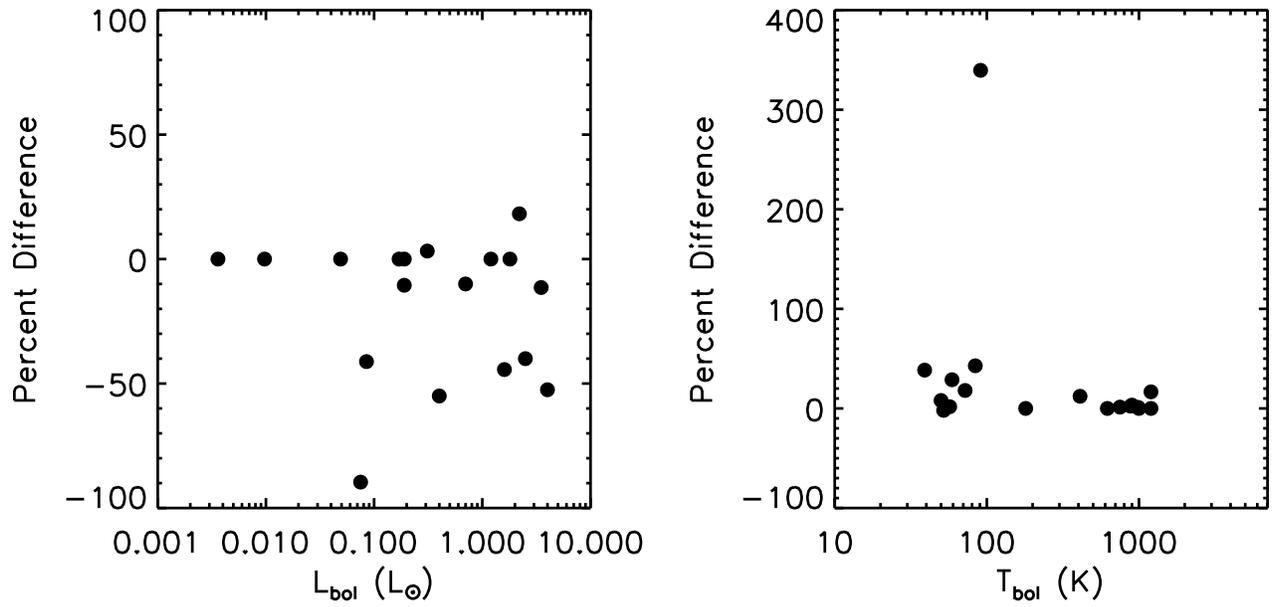}{7.0in}{0}{90}{90}{-300}{-200}
\caption{\label{fig_compare}\emph{Left:}  Percent difference between \lbol\ from Evans et al. (2008) and the new value of \lbol\ from this study.  \emph{Right:}  Percent difference between \tbol\ from Evans et al. (2008) and the new value of \tbol\ from this study.}
\end{figure}

To quantify the effects that our improved far-infrared and submillimeter 
photometry have on \lbol\ and \tbol, Figure \ref{fig_compare} shows, for both 
\lbol\ and \tbol, the percent difference for each source 
between the value calculated by Evans et al. (2008) and our value.  The 
results for both \lbol\ and \tbol\ are in good agreement with previous 
studies that find the error introduced in either quantity by integrating 
over incomplete, finitely-sampled SEDs is, on average, $20-60$\% 
(Enoch et al. 2008; Dunham et al. 2008).  The one source with a very large 
percent difference in \tbol\ is Source 75.  As noted in Table \ref{tab_lboltbol}, 
our value of \tbol\ $=91$ K is actually a lower limit, thus this 
large percent difference is an upper limit to the true percent difference.

We conclude that the combination of more accurate 70 $\mu$m photometry and 
adding sub-millimeter photometry at 350 $\mu$m does produce more accurate 
calculations of \lbol\ and \tbol, but the changes are generally not large 
enough to change source classifications (except for sources near the 
boundaries between classes) and are in agreement with previous studies.

\subsection{Comparison to Previous Studies}

In the region covered in our study, we find a total of 5 Class 0 sources, 
7 Class I sources, and 7 Class II/III sources.  
In a recent, multi-wavelength study of Cluster B, Djupvik et al. (2006) 
found a YSO population consisting of 2 Class 0 sources (one only 
tentatively suggested as Class 0), 5 Class I sources, 5 flat-spectrum 
sources, 31 Class II sources.  Their study covered a much larger area 
than our focused study on the cluster core and used a combination of
ISO mid-infrared data together with ground based near-infrared and
IRAM 1.3 mm data. Removing all sources from 
their sample not covered by our observations brings their sample size down to
2 Class 0 sources, 4 Class I sources, 1 flat-spectrum source, and 8 
Class II sources.  A natural question to ask is how well the two samples 
agree.

Of their 2 Class 0 sources, both are in our sample and also classified as 
Class 0.  Of their 4 Class I sources, all are in our sample.  Two are 
classified as Class 0, one as Class I, and one as Class II/III.  Their 
classification is based on the infrared spectral slope, which does not 
distinguish between Class 0 and Class I.  By sampling the full SED we are 
able to classify two sources (Sources 40 and 68) as Class 0 that can only be 
classified as Class I based on infrared data alone.  The disagreement in 
classification for the one source classified as Class II/III in our sample 
(Source 58) was discussed by Djupvik et al. (2006), who attributed the 
discrepancy to either strong H$_2$ line emission in their photometry or 
source variability.

Their one flat-spectrum source is in our sample as a Class I source (Source 
60).  There is no formal boundary in \tbol\ for flat-spectrum sources, 
although Evans et al. (2008) suggest a range of \tbol\ $=350-950$ K.  
This source, with \tbol\ $=620$ K, is within that range, thus our 
classifications of this source agree.

Of their 8 Class II sources, 7 are included in our sample.  Of these 7, all 
are classified as Class II/III except for Source 84, which we classify as 
Class I.  Djupvik et al. note that this is actually an unresolved binary, 
thus classification is difficult since we are attempting to classify the 
combined emission from two objects.  The remaining source is not classified 
as a YSO in the c2d survey.

Finally, there are five sources in our sample of YSOs that are not included 
in the Djupvik et al. (2006) sample:  Sources 42, 46, 60, B, and 75.  All 
but source 42 are too faint to be detected by Djupvik et al. (2006).  Source 
42 is a Class 0 object (\tbol\ $=52$ K) that may have been too deeply 
embedded to detect in their study.

In summary, our sample of Cluster B YSOs and the sample of YSOs presented 
by Djupvik et al. (2006) show good overlap.  Small discrepancies can 
be explained on a case-by-case basis, and we also find good agreement 
between source classifications, with a few discrepancies that likely 
result from our classification based on photometry that better samples 
the far-infrared and submillimeter peak of YSO SEDs.

\section{Summary}

We have obtained a significant improvement in the accuracy of 70\micron\
photometry with the Spitzer Space Telescope by utilizing the fine-scale
mode with MIPS at 70\micron\ for our observations of Cluster B.  The
improvements came jointly from the much improved sampling of the PSF
and the higher saturation limits at that spatial scale for the bright
objects in this cluster.  We have also been able to extend the SED's of
many of the YSO's in this cluster to longer wavelengths with the
addition of the SHARC-II 350\micron\ mapping.  The rough source classification
from the {\it c2d} project, however, has remained unchanged for most of
these objects, probably because at 24\micron\ their fluxes already gave
a reliable indication of their YSO classification.  Our observations of
VV Ser at 70 and 160\micron\ with much improved sampling have not revealed
any new structure or emission regions not seen in the earlier {\it c2d} studies.

\section{Acknowledgments}

Support for this work
was provided by NASA through RSA 1281173
 issued by the Jet Propulsion Laboratory, California Institute
of Technology, NASA Origins Grant NNX07AJ72G, and Spitzer contract 1288658, all
to the University of Texas at Austin.  We also benefitted greatly from comments
on earlier drafts of this paper by Neal Evans II, Karl Stapelfeldt, and Yancy
Shirley.

\clearpage

\begin{table}[h]
\caption{Observations Summary (Program ID = 20063) \label{aortable}}
\vspace {3mm}
\begin{tabular}{lccc}
\tableline
\tableline
AOR  & Date & Wavelengths & BCD Process \cr
\tableline

\bf{Cluster B Observations}\cr
%\bigskip

\dataset{ads/sa.spitzer\#0016795904} & 2006-10-04 & 70\micron-fine & S14.4.0 \\
\dataset{ads/sa.spitzer\#0016796160} & 2006-05-05 & 70\micron-fine & S14.4.0 \\
\dataset{ads/sa.spitzer\#0016796416} & 2006-09-30 & 70\micron-fine & S14.4.0 \\
\dataset{ads/sa.spitzer\#0016796672} & 2006-05-05 & 70\micron-fine & S14.4.0 \\
\tableline

\bf{VV Ser Observations}\cr
%\bigskip

\dataset{ads/sa.spitzer\#0016796928} & 2007-05-21 & 70\micron-fine, 160\micron\ & S16.1.0 \\
\dataset{ads/sa.spitzer\#0016797184} & 2007-05-20 & 70\micron-fine, 160\micron\ & S16.1.0 \\
\dataset{ads/sa.spitzer\#0016797440} & 2007-05-20 & 70\micron-fine, 160\micron\ & S16.1.0 \\

\tableline
\end{tabular}
\end{table}

\begin{deluxetable}{cccccccccc}
\tabletypesize{\small}
\rotate
\tablecolumns{8}
\tablecaption{Cluster B Young Objects-c2d Fluxes and This Paper\label{ysotable}}
\tablewidth{0pc}
\tablehead{
\colhead{YSO}                &
\colhead{Name/Position} &
%\colhead{Prev. Name\tablenotemark{a}} &
\colhead{3.6 \micron\ }  &
\colhead{4.5 \micron\ }  &
\colhead{5.8 \micron\ }  &
\colhead{8.0 \micron\ }  &
\colhead{24.0 \micron\ }  &
\colhead{70.0 \micron\ {\it c2d}}  &
\colhead{70.0 \micron\ } &
\colhead{350 \micron\ }          \\
\colhead{\#}                &
\colhead{SSTc2dJ...} &
%\colhead{}                &
\colhead{(mJy)}  &
\colhead{(mJy)}  &
\colhead{(mJy)}  &
\colhead{(mJy)}  &
\colhead{(mJy)}  &
\colhead{(mJy)}  &
\colhead{(mJy)}  &
\colhead{(Jy)}
}

\startdata

 37 & 18285276$+$0028467 & 1.84$\pm$0.10 & 2.45$\pm$0.14 & 2.58$\pm$0.15 & 3.44$\pm$0.19 & 15.7$\pm$ 1.5 & \nodata & $<$ 84  & $<$0.6 \\
 40 & 18285404$+$0029299 & 5.81$\pm$0.50 & 27.6$\pm$ 2.3 & 44.8$\pm$ 2.6 & 56.4$\pm$ 3.2 &  918$\pm$  85 & 11100$\pm$ 1040 & 15260$\pm$3050  & 9.8$\pm$2.0 \\
 41 & 18285450$+$0028523 & 14.7$\pm$ 0.9 & 34.2$\pm$ 2.0 & 44.8$\pm$ 2.3 & 25.4$\pm$ 1.4 & 4.53$\pm$0.48 & \nodata & $<$55  & $<$0.6 \\
 42 & 18285486$+$0029525 & 1.94$\pm$0.12 & 10.6$\pm$ 0.6 & 20.4$\pm$ 1.1 & 30.2$\pm$ 1.6 &  765$\pm$  70 & 7250$\pm$  675 & 4840$\pm$970  & 4.7$\pm$1.0 \\
 45 & 18285577$+$0029447 & 0.26$\pm$0.02 & 1.87$\pm$0.14 & 2.23$\pm$0.14 & 3.08$\pm$0.17 &  126$\pm$  11 & \nodata & 1470$\pm$295  & 4.0$\pm$0.8 \\
 46 & 18285664$+$0030082 &0.055$\pm$0.007 & 0.14$\pm$0.01 & 0.12$\pm$0.03 & 0.22$\pm$0.05 & 13.0$\pm$ 1.2 & \nodata & $<$110  &  $<$0.6 \\
 50 & 18285945$+$0030031 & 38.4$\pm$ 2.1 & 41.0$\pm$ 2.2 & 43.5$\pm$ 2.3 & 49.4$\pm$ 2.7 & 81.6$\pm$ 7.6 &  204$\pm$  32 & 238$\pm$48  & $<$0.6 \\
 54 & 18290089$+$0029316 &  246$\pm$  13 &  290$\pm$  16 &  308$\pm$  19 &  392$\pm$  23 &  711$\pm$  67 &  736$\pm$  75 & 1080$\pm$216  & $<$0.6 \\
 55 & 18290107$+$0031452 & 59.2$\pm$ 3.6 & 72.8$\pm$ 4.3 & 76.2$\pm$ 4.1 & 75.5$\pm$ 4.3 & 72.5$\pm$ 6.7 & \nodata & $<$100  & $<$0.6\\
 56 & 18290122$+$0029330 & 88.8$\pm$ 4.8 & 97.4$\pm$ 5.1 & 91.0$\pm$ 5.3 &  100$\pm$   6 &  215$\pm$  21 & \nodata & $<$300  & $<$0.6\\
 58 & 18290175$+$0029465 &  141$\pm$   8 &  133$\pm$   6 &  111$\pm$   6 &  107$\pm$  10 &  361$\pm$  33 & \nodata & 467$\pm$94  & $<$0.6 \\
 59 & 18290184$+$0029546 &  586$\pm$  51 &  553$\pm$  33 &  504$\pm$  28 &  461$\pm$  27 &  407$\pm$  38 &  503$\pm$  52 & 430$\pm$86  & $<$0.6 \\
 60 & 18290211$+$0031206 & 1.19$\pm$0.07 & 1.62$\pm$0.09 & 1.58$\pm$0.10 & 1.13$\pm$0.07 & 22.1$\pm$ 2.0 &  276$\pm$  29 & 536$\pm$107  & $<$0.6 \\
 61 & 18290283$+$0030095 & 15.4$\pm$ 1.0 & 19.2$\pm$ 1.1 & 34.5$\pm$ 2.0 & 30.6$\pm$ 1.8 & 94.2$\pm$ 8.7 &  535$\pm$  54 & 700$\pm$140  & $<$0.6 \\
 67 & 18290619$+$0030432 & 8.05$\pm$0.41 & 45.0$\pm$ 2.8 & 93.9$\pm$ 4.8 &  129$\pm$   7 & 1320$\pm$  139 & 7240$\pm$  713 & 11150$\pm$2230  & 13$\pm$6 \\
 68 & 18290675$+$0030343 & 3.27$\pm$0.21 & 11.7$\pm$ 0.7 & 14.9$\pm$ 0.8 & 20.7$\pm$ 1.2 & 1000$\pm$  105 & 11400$\pm$  1180 & 25305$\pm$5060  &  13$\pm$6\\
 B\tablenotemark{a}  & 18290864$+$0031305 & 0.06$\pm$0.03 & 0.32$\pm$0.02 & 0.47$\pm$0.05 & 0.62$\pm$0.07 & 36.2$\pm$ 3.4 & \nodata & 71$\pm$110\tablenotemark{b} & $<$0.6 \\
 75 & 18290904$+$0031280 & 0.95$\pm$0.11 & 2.78$\pm$0.23 & 2.92$\pm$0.24 & 5.03$\pm$0.40 & 14.0$\pm$ 1.9 & \nodata & 766$\pm$490\tablenotemark{b}  & $<$3.0\\
 C\tablenotemark{a}  & 18290906$+$0031323 & $<$  0.12     & 0.29$\pm$0.03 & 0.40$\pm$0.09 & 0.31$\pm$0.08 & 64.6$\pm$ 6.0 & 6380$\pm$  638  & 7905$\pm$1580   & 13.5$\pm$2.8 \\

\enddata
\tablenotetext{a}{Source designation from Table 7 in \citet{harv07a}.}
\tablenotetext{b}{Formal fluxes and uncertainties from the model fitting.  These clearly
should be viewed as upper limits of three times the uncertainties.}
\end{deluxetable}

\begin{deluxetable}{lrrcrr}
\tabletypesize{\scriptsize}
\tablewidth{0pt}
\tablecaption{\label{tab_lboltbol}Bolometric Luminosities and Temperatures of Cluster B YSOs}
\tablehead{
\colhead{YSO} & \colhead{\lbol} & \colhead{\tbol} & & \colhead{c2d \lbol\tablenotemark{b}} & \colhead{c2d \tbol\tablenotemark{b}} \\
\colhead{\#}   & \colhead{(\lsun)} & \colhead{(K)} & Classification\tablenotemark{a} & \colhead{(\lsun)} & \colhead{(K)}}
\startdata
37 & 0.0097 & 620 & I & 0.0097 & 620 \\
40 & 3.5 & 57 & 0 & 3.1&  58 \\
41 & 0.049 & 750 & II/III & 0.049 & 760 \\
42 & 2.2 & 52 & 0 & 2.6 & 51 \\
45 & 0.40 & 50 & 0 & 0.18 & 54 \\
46 & 0.0036 & 180 & I & 0.0036 & 180 \\
50 & 0.17 & 880 & II/III & 0.17 & 900 \\
54 & 1.2 & 900 & II/III & 1.2 & 930 \\
55 & 0.19 & 990 & II/III & 0.19 & 1000 \\
56 & 0.31 & 1000 & II/III & 0.32 & 1000 \\
58 & 0.70 & 1200 & II/III & 0.63 & 1400 \\
59 & 1.8 & 1200 & II/III & 1.8 & 1200 \\
60 & 0.085 & 72 & I & 0.050 & 85 \\
61 & 0.19 & 410 & I & 0.17 & 460 \\
67 & 2.5 & 84 & I & 1.5 & 120 \\
68 & 4.0 & 59 & 0 & 1.9 & 76 \\
B & $<$0.019 & $>$140 & I & ...&  ... \\
75 & $<$0.075& $>$91 & I & 0.0078 & 400 \\
C & 1.6 & 39 & 0 & 0.89&  54 \\
\enddata \\
\tablenotetext{a}{Classification based on \tbol; see text for details.}
\tablenotetext{b}{Values of \lbol\ and \tbol\ taken from Evans et al. (2008).}
\end{deluxetable}

%%%%%%%%%%%%%%%%%% Tables %%%%%%%%%%%%%%%%%%%%%

%\clearpage
%%%%%%%%%%%%%%%%%% Figures %%%%%%%%%%%%%%%%%%%%%

\clearpage

\end{document}